\documentclass[aps,prb,twocolumn]{revtex4}
\usepackage{graphicx}
\usepackage{amssymb}

\begin{document}

\title{The Quantum Capacitance Detector: A concept for a pair-breaking radiation detector based on the single Cooper-pair box}

\author{M. D. Shaw,$^1$ J. Bueno,$^2$ P. Day,$^2$, C. M. Bradford$^2$, and P. M. Echternach,$^2$}

\address{$^1$ Department of Physics and Astronomy, University of Southern California, Los Angeles, California 90089, USA}
\address{$^2$ Jet Propulsion Laboratory, California Institute of Technology, Pasadena, California 91109, USA}

\date{\today}

\begin{abstract}
We present a proposed design for a pair-breaking photodetector for far-infrared and sub-millimeter radiation. Antenna-coupled radiation generates quasiparticles in a superconducting absorber, the density of which are measured using a single Cooper-pair box. Readout is performed using an electromagnetic oscillator or a microwave resonator, which is well suited for frequency multiplexing in large arrays. Theoretical limits to detector sensitivity are discussed and modeled, with predicted sensitivities on the order of $10^{-21}~\mathrm{W}/\sqrt{\mathrm{Hz}}$. We anticipate that this detector can be used to address key scientific goals in far-infrared and sub-millimeter astronomy. 
\end{abstract}

\maketitle

	Cryogenic radiation detectors based on low-temperature superconductors have a long and successful history in a wide variety of applications, including astronomy,\cite{Peacock} high-energy physics,\cite{particles} optical science,\cite{saewoo} and communications\cite{nanowire} over a spectral range from x-rays to the sub-millimeter. Various physical processes have been exploited to create sensitive detectors, which can be roughly separated into bolometric devices, such as the transition-edge sensor\cite{firstTES} and the hot-electron bolometer mixer,\cite{firstHEB,nanoHEB} and pair-breaking devices, such as the superconducting tunnel junction (STJ) detector\cite{Peacock,firstSTJ} and the microwave kinetic inductance detector (MKID) .\cite{MKID} In a pair-breaking detector, radiation coupled to a superconducting absorber breaks Cooper pairs in the material, generating quasiparticles. A key challenge in designing pair-breaking detectors is the readout, the means of measuring the number of quasiparticles in the absorber in the presence of the Cooper pairs. This has been accomplished by measuring a current across a tunnel junction, as in the STJ, and by measuring a change in the impedance of a superconducting resonator, as in the MKID. 
	
	In this work, we propose a pair-breaking detector appropriate for sub-millimeter radiation based on the single Cooper-pair box (SCB), a mesoscopic superconducting circuit which has been extensively studied in the context of quantum computation.\cite{schonRMP} The SCB has also been shown to be extremely sensitive to the presence of non-equilibrium quasiparticles.\cite{Aumentado1,Aumentado2,Ferguson,ourpaper} The proposed device, which we call the quantum capacitance detector (QCD), uses the SCB to sample the density of quasiparticles in the absorber. The QCD	promises excellent sensitivity, with estimates of the minimum noise-equivalent power (NEP) on the order of $10^{-21}~\mathrm{W}/\sqrt{\mathrm{Hz}}$ under realistic operating conditions. Like the MKID, the QCD can be read out in the frequency domain, naturally lending itself to large-scale multiplexing using well-established fabrication and measurement techniques. Among many other potential applications, large arrays of sensitive low-noise detectors are critically needed to meet scientific goals in far-infrared and sub-millimeter astronomy.\cite{BLISS}

\begin{figure}[t]
\centering
\includegraphics{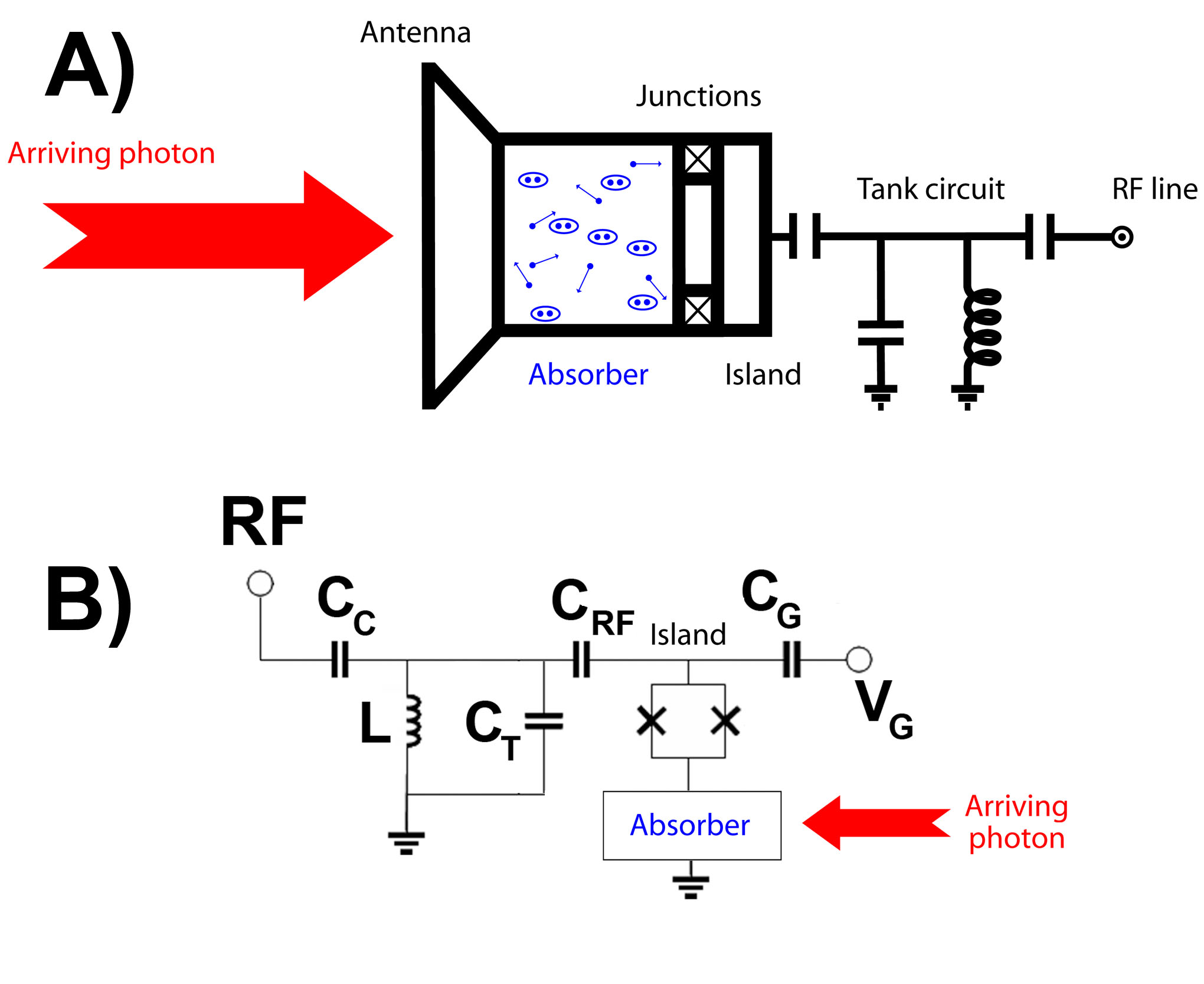}
\caption{(Color online) \it A)\rm~Cartoon schematic of the QCD concept. Incident submillimeter radiation is coupled by an antenna into a superconducting absorber, generating quasiparticles. The quasiparticle density in the absorber is then measured with an SCB. \it B)\rm~Circuit diagram of the SCB and its readout.}
\end{figure}

\section{Detection Concept}

\subsection{Overview}

The SCB consists of a small superconducting island coupled to superconducting leads (hereafter referred to as the \it absorber\rm) by a pair of ultra-small Josephson junctions, typically arranged in a DC-SQUID configuration. A schematic of the SCB is shown in figure 1. The capacitance of the island, which is typically on the order of a few fF, is small enough that the charging energy is the dominant term in the Hamiltonian, and charge number states are well defined. As the potential of the island is adjusted with a gate capacitor, the two lowest energy levels form an avoided level crossing. At the charge degeneracy point, where the lowest-lying eigenstates are symmetric and antisymmetric superpositions of charge states, the SCB may be treated as a two-level quantum system coupled to an effective magnetic field. In the QCD concept, the state of the SCB is read out dispersively using a quantum capacitance technique,\cite{DelsingQCR,FinnQCR} where the SCB island is coupled capacitively to an LC oscillator which is slow compared to the qubit energy level splitting. By probing the oscillator with RF reflectometry, one can directly measure the quantum capacitance of the SCB, which is proportional to the second derivative of the SCB energy level with respect to the gate charge. An equivalent measurement can also be performed using an inductive coupling between the SCB and the oscillator.\cite{BornQCR} Note that in the QCD scheme, the SCB is always operated in its ground state, obviating the need for coherent manipulation. 

Although experiments aimed toward quantum computation involve delicate superpositions of Cooper pairs, the SCB is exquisitely sensitive to incoherent quasiparticle tunneling between the absorber and the island, even at temperatures where thermally excited quasiparticles are strongly frozen out. When non-equilibrium quasiparticles tunnel across the SCB junctions, the parity of the device abruptly switches, destroying quantum coherence. As such, this effect is typically known as quasiparticle ``poisoning". However, this property can be exploited to make an extremely sensitive measurement of the quasiparticle density in the absorber, since the quasiparticle tunneling rates and the fraction of time the device spends in each parity state are related in an extremely simple way to the quasiparticle density in the absorber. 

\begin{figure}[t]
\centering
\includegraphics{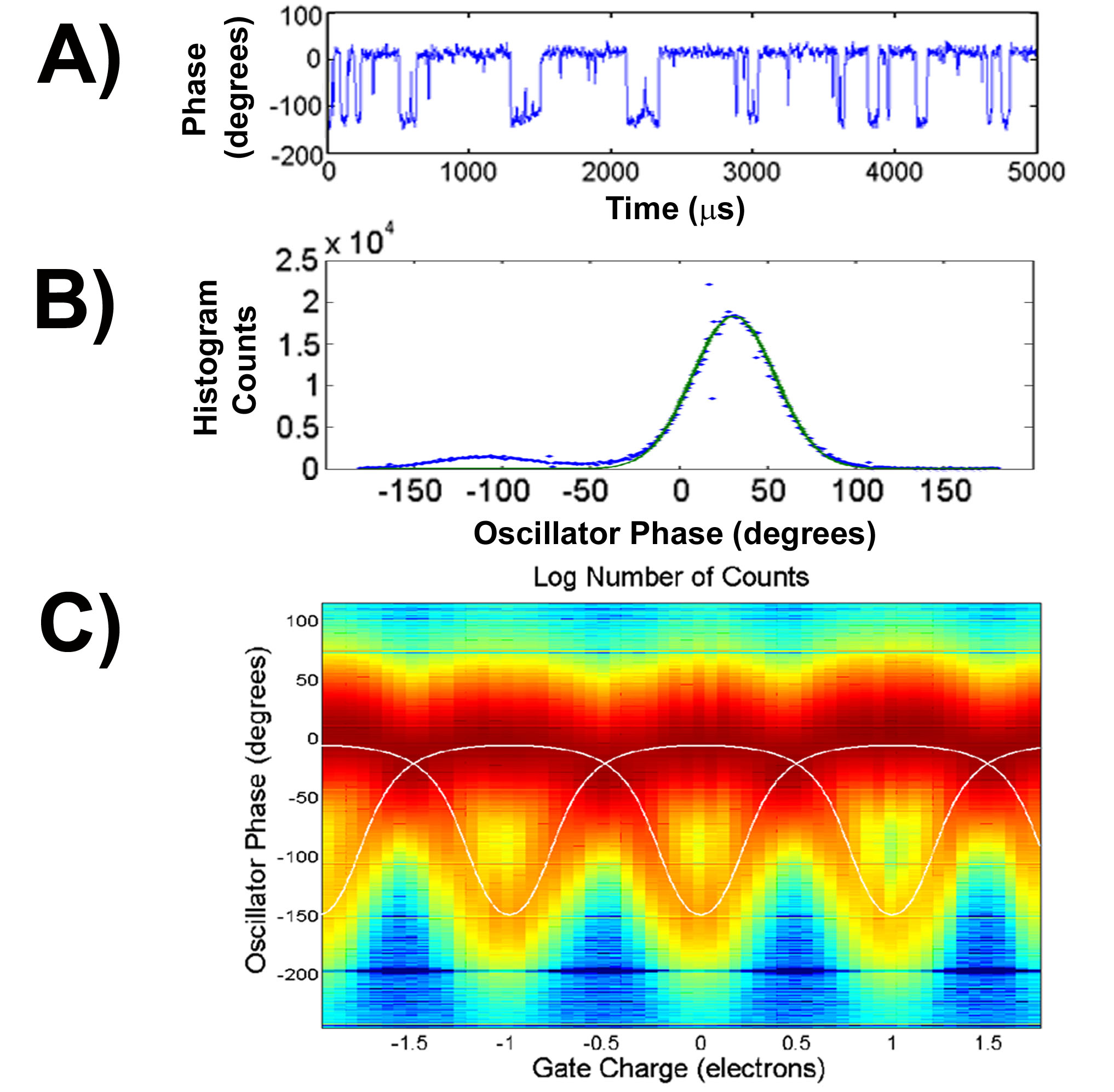}
\caption{(Color online) \it A)\rm~Example of two-rate telegraph noise in the capacitance signal, recorded at the SCB degeneracy point. Data adapted from Ref.~(14). \it B)\rm~Phase shift histogram, taken at the degeneracy point, showing two clearly separated peaks corresponding to odd and even parity states. Note the large ($140^\circ$) phase shift between the two. The phase shift signal in the QCD is an average taken over the entire time signal, as indicated by the red line. \it C)\rm~Phase shift histograms as a function of gate voltage. The z-axis is plotted logarithmically to emphasize the presence of the minority (odd) parity state. The white lines are theoretically predicted capacitance traces for the odd and even parity states.}
\end{figure}

	The architecture of the QCD is shown in figure 1. Infrared or sub-millimeter radiation coupled to the antenna provides energy to break Cooper pairs in the absorber, generating phonons and nonequilibrium quasiparticles which diffuse to the junctions of the SCB. In aluminum, an attractive material for fabrication of SCBs and superconducting resonators, pair-breaking can occur for photons with a frequency above 90 GHz. The SCB is capacitively coupled to an LC oscillator or microwave resonator, which is used to measure the SCB state. 
	
	When a single quasiparticle tunnels onto the SCB island, it switches the parity, changing the effective gate charge by one electron, bringing the SCB far from its degeneracy point and changing its capacitance. This sudden shift in the capacitance leads to a large frequency shift in the oscillator, which was measured to be $140^\circ$ in a recent experiment,\cite{ourpaper} and is shown in Fig.~2. When quasiparticle tunnels back into the absorber, the parity switches back and the device returns to the degeneracy point. This results in a time trace given by a two-rate random telegraph signal. By measuring the average center frequency of the oscillator, one can average over this telegraph signal and extract the fraction of the time the SCB spends in the ``even" and ``odd" parity states, which is simply related to the tunneling rates and hence the density of quasiparticles in the absorber. In this paper, we refer to the even parity state as the one with zero or an even number of quasiparticles on the island, and the odd state as the one with one or an odd number of quasiparticles on the island. 
	
	In the QCD concept, the absorber is small compared to the characteristic diffusion length of quasiparticles in Al, so that the spatial distribution of quasiparticles in the absorber may be taken to be uniform, and any given quasiparticle may be assumed to rapidly sample the entire volume of the absorber. Quasiparticles generated in the higher-$T_C$ antenna will be trapped in the absorber by Andreev reflection. A key advantage of the SCB as a detector readout is that it is sensitive to the density of quasiparticles, rather than the overall number. Thus the absorber volume is a key design parameter which can be used to tune the operating wavelength range, sensitivity, and saturation power, as discussed below.

	The QCD concept is related to but clearly distinct from previous proposals to use single-charge devices for electromagnetic detection. In one such experiment involving a single-electron transistor (SET) with a superconducting island and normal leads, Andreev-cycle transport was switched on and off when a quasiparticle from the leads was transferred to the island via photon-assisted tunneling.\cite{tinkham} In a second experiment, quasiparticle tunneling current in a biased submicron SIS junction is measured using an RF-SET.\cite{schoelkopf_detector} However, detectors based on the superconducting SET present significant challenges for multiplexed readout, due to the need to transform the high impedance of the SET junctions to that of a $50~\Omega$ transmission line. 

\subsection{Strategies for Multiplexed Arrays}
	
Another significant advantage of the QCD concept is the natural use of frequency multiplexing to read many pixels simultaneously with a single high-frequency line.  Each pixel in a QCD array would have a different resonant frequency, determined by the values of the on-chip inductor and capacitor or the length of the resonator.  All pixels could be read out simultaneously by applying a frequency comb to all devices through a common transmission line.  Each frequency in the comb would match that of a single resonant circuit, and the reflected power at that frequency would constitute the SCB readout signal for that pixel. Signal generation, demodulation, and analysis for large arrays can be performed using technology developed for software defined radio, and such measurement techniques are already in development for multiplexed MKID arrays.\cite{MKIDcamera} A design for the electronics required to implement such a multiplexing scheme is shown in Fig.~3. A simple multiplexed quantum capacitance readout of two SCBs has been recently demonstrated.\cite{ourMultiplexed}

\begin{figure}[t]
\centering
\includegraphics{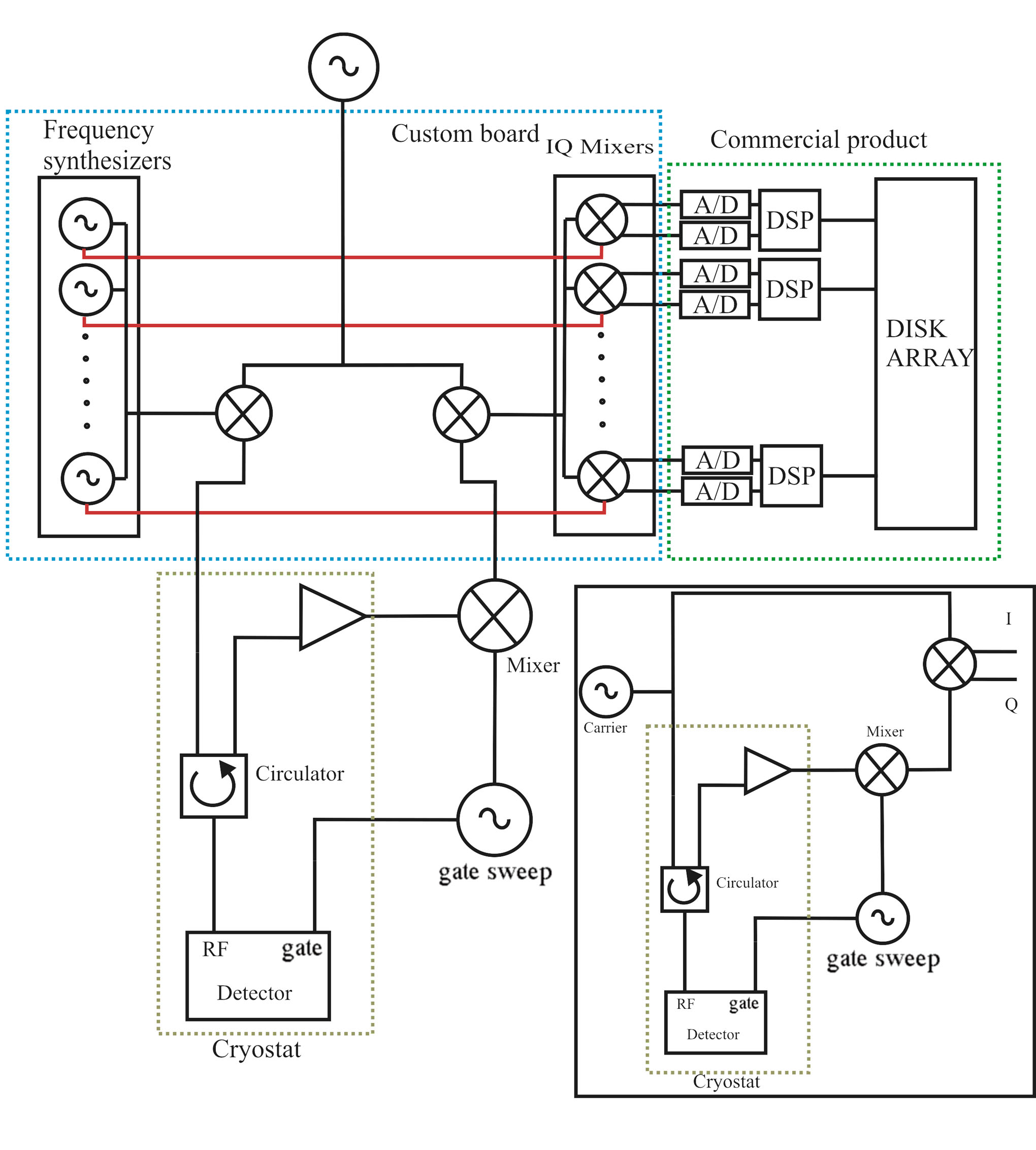}
\caption{(Color online) Schematic block diagram of proposed QCD readout electronics. \it Inset, lower right\rm: Single pixel - An RF carrier signal excites the tank circuit of a particular detector pixel, producing a reflected signal that is phase shifted according the average charge state of the SCB.  An AC bias voltage is applied, with an amplitude of $e/C_g$, which modulates the phase shift signal by sweeping the SCB through one full gate period.  A mixer demodulates the reflected RF signal at the bias voltage modulation frequency, and a down-converter translates the result to DC.  \it Main\rm: Multiplexed readout - A low frequency comb function (0-200 MHz) containing several frequency components is produced digitally and block up-converted, resulting in a comb of RF carrier frequencies, each corresponding to a particular detector.  All of the SCB gate lines are tied together and modulated at the same frequency.}
\end{figure}

Since the operating point of the SCB is sensitive to small static charge inhomogeneities, it is impractical to tune the operating points of all SCB pixels in a QCD array to the degeneracy point with a single DC gate line. As shown in Fig.~2, the response of the SCB to parity fluctuations is maximal at the degeneracy point. However, this problem can be readily solved by averaging over all gate voltages, as shown schematically in Fig.~3. An AC voltage with an amplitude of $e/C_g$ (where $C_g$ is the gate capacitance) can be applied to all SCB gates through a common line. This adiabatically sweeps the operating point of each SCB through one full period, modulating the RF output signal.  A mixer demodulates the reflected RF signal at the AC modulation frequency, and a down-converter translates the result to DC.  In this way, a single AC tone applied to all SCB gates will appear as a sideband of each RF readout frequency in the comb, which gives the time-averaged capacitance averaged over all gate voltages when mixed down to DC. The reflected RF comb, containing the phase shift information for the entire array, can be demodulated at the gate modulation frequency, down-converted to the 0-200 MHz band, then digitized and digitally demultiplexed.

\section{Device Physics}

\subsection{Quasiparticle Tunneling}

	The physics of nonequilibrium quasiparticle tunneling in single Cooper-pair devices has been extensively studied experimentally and can be effectively described by a kinetic trapping theory.\cite{Lutchyn} With the SCB biased at the degeneracy point, the island will behave as a potential well or ``trap" for nonequilibrium quasiparticles in the absorber, with an effective depth $\delta E \cong E_C - \frac{E_J}{2} - \tilde{\Delta}$, where $E_C = \frac{e^2}{2C_\Sigma}$ is the single-electron charging energy, $C_\Sigma$ is the overall capacitance of the SCB island, $E_J$ is the Josephson energy, which is proportional to the critical current of the junctions, and $\tilde{\Delta} = \Delta_I - \Delta_A$ is the island-absorber superconducting gap profile. While in superconducting circuits for quantum computation $\tilde{\Delta}$ is frequently engineered to be as large as possible to suppress quasiparticle tunneling,\cite{Aumentado1} in the QCD it is advantageous to have $\tilde{\Delta} \leq 0$. For the remainder of this paper, we assume that $\tilde{\Delta} = 0$. 
	
	Given a uniform nonequilibrium quasiparticle density $n_{qp}$ in the leads of the SCB, quasiparticles will tunnel onto the island at a proportional rate $\Gamma_{\mathrm{in}} = Kn_{qp}$. The proportionality constant is given by
	
\begin{equation}
\label{Kfactor}
K = \frac{G_N}{e^2}\frac{e^{\Delta/k_BT}}{N_L} \int^\infty_\Delta
dE~h(E) e^{-E/k_BT}
\end{equation}

\noindent where $N_L = D(E_F)\sqrt{2\pi\Delta k_BT}$ is the density of quasiparticle states available in the lead, $D(E_F)$ is the normal-metal density of states at the Fermi level, $G_N$ is the tunneling conductance, $T$ is the sample temperature, and $k_B$ is the Boltzmann constant. Furthermore, the effective density of states is given by

\begin{equation}
\label{DOS}
h(E) = \frac{E(E+\delta E) - \Delta^2}{\sqrt{\left((E+\delta E)^2 - \Delta^2\right) \left(E^2 - \Delta^2\right)}}
\end{equation}

\noindent which includes BCS coherence factors accounting for quantum interference between electron-like and hole-like quasiparticle tunneling.\cite{Lutchyn} The approximation of a tunneling rate proportional to the quasiparticle density is valid as long as the operation temperature $k_B T \ll \Delta$. 

To a first approximation, the tunnel rate $\Gamma_{\mathrm{out}}$ for nonequilibrium quasiparticles on the SCB island back onto the absorber is independent of the quasiparticle density $n_{qp}$. The probability that the SCB will be found in the odd parity state then depends on the quasiparticle density in a particularly simple way,

\begin{equation}
\label{Peven}
P_{\mathrm{odd}}(n_{qp}) = \frac{1}{1 + \Gamma_{\mathrm{out}}/Kn_{qp}}.
\end{equation}

\noindent A theoretical expression for $\Gamma_{\mathrm{out}}$, which involves two separate time scales for elastic and inelastic quasiparticle tunneling, can be found in Ref.~(\cite{Lutchyn}). In the proposed detector, $P_{\mathrm{odd}}$ is the experimentally accessible quantity which is used to measure the quasiparticle density. At 100 mK, the proposed QCD operating temperature, both $K$ and $\Gamma_\mathrm{out}$ are only weakly dependent on temperature.\cite{ourpaper} They are primarily set by the material and qubit parameters. 

\subsection{Oscillator Response}

When a quasiparticle generated by incident radiation tunnels onto the SCB island, the SCB parity switches, changing the quantum capacitance $C_Q$. This shift in capacitance leads to a shift in the center frequency of the oscillator to which the SCB is capacitively coupled. Within the bandwidth of the oscillator, these shifts can be observed directly in the time domain to obtain the tunnel rates and hence the quasiparticle density,\cite{Aumentado2,Ferguson} but in a functioning detector it is more practical to use the time-averaged phase shift of the RF carrier reflected by the oscillator to measure the quasiparticle density and detect an incoming photon.  

In a parallel-element tank circuit with inductance $L$ and capacitance $C$ coupled to a transmission line through a capacitance $C_C$, as shown in Fig.~1, the phase of a reflected wave is given by $\tan\phi = -2|Z|Z_0/(|Z|^2 - Z_0^2)$ where 

\begin{equation}
\label{impedance}
Z = \frac{1-\left(\omega/\omega_0\right)^2-\left(\omega/\omega_C\right)^2}{i\omega C_C \left[1-\left(\omega/\omega_0\right)^2\right]}
\end{equation}

\noindent is the impedance of the tank circuit and $Z_0 \approx 50~\Omega$ is the impedance of the transmission line. In this expression, $\omega$ is the angular drive frequency, $\omega_0 = 1/\sqrt{L(C+C_Q)}$ is the center frequency, and $\omega_C = 1/\sqrt{LC_C}$. Modulation of the phase shift with the quantum capacitance comes in through modulation of $\omega_0$ in Eq.~(\ref{impedance}). 

The quantum capacitance of the SCB is proportional to the second derivative of the SCB energy with respect to the gate charge. When the SCB is tuned to its degeneracy point in the even state, the change in capacitance when a single quasiparticle tunnels is given by 

\begin{equation}
\label{capacitance}
\delta C_Q = \frac{C_g^2}{C_\Sigma} \frac{4E_C}{E_J} \left( 1 - \left[1+\left(\frac{4E_C}{E_J}\right)^2\right]^{-3/2} \right) 
\approx \frac{C_g^2}{C_\Sigma} \frac{4E_C}{E_J}
\end{equation}

\noindent for an SCB in the charge-eigenstate limit $E_C \gg E_J$. In this expression, $C_g$ is the coupling capacitance between the oscillator and the SCB island, while $C_\Sigma$ is the total capacitance of the SCB. Ideally, one would like to design the SCB and oscillator parameters so that the phase shift $\delta\phi$ induced by $\delta C_Q$ is $180^\circ$. In the test device discussed below, where the oscillator $Q \approx 3000$, the observed phase difference between the even and odd states was approximately $\delta\phi = 140^\circ$, as can be seen in Fig.~2. 

When performing reflectometry with a high-Q oscillator, the observed quantity is the time-averaged phase shift $\langle \phi \rangle$. The oscillator itself performs a moving average with a window length equal to the inverse of its bandwidth, and further averaging can be performed in software or with an external circuit. At all but the lowest temperatures and shortest time scales, the phase shift switches stochastically between its even- and odd-state values in a homogeneous two-rate Poisson random telegraph process with additive Gaussian noise. Since such a process is ergodic, we may treat the observed phase shift as the ensemble average

\begin{equation}
\label{phase_average}
\langle \phi \rangle = P_{\mathrm{odd}} \phi_{\mathrm{odd}} + \left(1-P_{\mathrm{odd}}\right) \phi_{\mathrm{even}}
\end{equation}

\noindent where $\phi_{\mathrm{even,odd}}$ are the phase shifts in the even and odd SCB parity states, and the deviation of the observed phase shift from its ``dark" value is $\phi = \langle\phi\rangle - \phi_{\mathrm{even}} = P_{\mathrm{odd}} \delta\phi$. In this way it is quite straightforward to extract the quasiparticle density from the phase shift of the reflected wave.

\subsection{Dark Counts}

In the development of single Cooper-pair devices for quantum computation, quasiparticle tunneling has generally been regarded as an unwanted effect which is a challenge to reproducibly control. As a result, it is reasonable to expect that in QCD devices there will be significant quasiparticle tunneling even when no signal photons are present. In applications where it is necessary to control the number of dark counts, quasiparticle tunneling can be suppressed by creating an energetic barrier for quasiparticle tunneling between the absorber and the island. This is typically realized in practice by creating a gradient in the superconducting gap energy between the absorber and the island, either by controlling impurity concentrations\cite{Aumentado1} or film thicknesses.\cite{nakamura_filmthickness,ferguson_filmthickness} However, the use of such techniques are likely to limit the sensitivity of the detector, requiring a careful engineering tradeoff. Of course, it is also possible that the number of dark counts can be reduced significantly through improved control over materials and more careful electromagnetic filtering and shielding. 

\section{Noise Sources and Detector Sensitivity}

\subsection{Sensitivity}

Given that single-charge devices such as the SCB are exquisitely sensitive to the presence of non-equilibrium quasiparticles, one of the key advantages of the QCD scheme is its extreme sensitivity to electromagnetic radiation at submillimeter wavelengths. For a material with superconducting gap energy $\Delta$, an incident photon with energy $h\nu$ will generate a number of quasiparticles $N_{\mathrm{qp}} = \eta h \nu / \Delta$ in the absorber, where the factor $\eta \approx 0.57$ is the efficiency with which the energy of the initial photoelectron is downconverted into quasiparticles.\cite{QPdownconversion} Assuming that the distribution of quasiparticles in the absorber is spatially uniform (that is, that the dimensions of the absorber are small compared to the quasiparticle diffusion length) this will produce a non-equilibrium quasiparticle density $n_{\mathrm{qp}} = N_{\mathrm{qp}} / \Omega$ where $\Omega$ is the absorber volume. 

The sensitivity of the QCD system can be quantified by the noise-equivalent power (NEP), which is defined as the radiant power per square root bandwidth required to achieve a signal-to-noise ratio (SNR) of unity at a given signal modulation frequency. Since the final signal in the QCD is the mean phase shift in an oscillator, all relevant sources of noise can be expressed as phase noise. For a phase noise spectral density $S_\phi (\omega)$, the NEP is given by 

\begin{equation}\label{NEPgeneral}
\mathrm{NEP}(\omega) = \frac{dP_s}{dn_{qp}}\left( \frac{d\phi}{dn_{qp}} \right)^{-1} 
\sqrt{ S_\phi (\omega) \left( 1 + \omega^2 \tau_{qp}^2 \right) 
\left( 1 + \omega^2 \tau_{r}^2 \right)}
\end{equation} 

\noindent where $\tau_{qp}$ is the quasiparticle lifetime, $P_s$ is the incident radiant source power, and $\tau_r$ is the ringdown time of the oscillator. In this equation, the responsivity 

\begin{equation}
\label{responsivity}
\frac{d \phi}{dn_{\mathrm{qp}}} =
\frac{\delta\phi\Gamma_{\mathrm{out}} K}{\left(Kn_{\mathrm{qp}} + \Gamma_{\mathrm{out}}\right)^2}.
\end{equation}

\noindent is simply defined as the overall average phase shift per quasiparticle, as in the MKID.\cite{MKID} 

The source power is related to the quasiparticle density through the rate equation

\begin{equation}
\frac{dn_{qp}}{dt} = \frac{\eta P_s}{\Delta \Omega} - \frac{n_{qp}}{\tau_{qp}} - R n_{qp}^2
\end{equation}

\noindent where $R$ is the recombination constant. In aluminum, $R = 9.6~\mu\mathrm{m}^3/\mathrm{s}$.\cite{Gaidis}  The steady-state solution is given by 

\begin{equation}\label{Nqp}
n_{qp} = \frac{1}{2R\tau_{qp}} \left( \sqrt{ 1 + \frac{4\eta R P_s \tau_{qp}^2}{\Delta \Omega}} - 1 \right)
\approx \frac{\eta P_s \tau_{qp}}{\Delta\Omega}.
\end{equation}

\noindent The phase noise in the QCD signal arises from a variety of physical mechanisms. The contributions to the NEP from the dominant sources of noise are discussed below. 

\subsection{Telegraph Noise}

In the QCD scheme, the quasiparticle density in the absorber is sampled by measuring the rates of quasiparticle tunneling in the SCB, which occurs in discrete events. In practice, the observed quantities are the occupation probabilities in the even and odd parity states, which are extracted from the average phase shift of the oscillator. Since this average must be performed in time over the random telegraph signal intrinsic to quasiparticle tunneling, a significant source of phase noise in the QCD is ``telegraph noise", or quasiparticle shot noise. 

The telegraph noise in the QCD can be characterized as a two-state random process with two rates, corresponding physically to tunneling from the absorber to the island ($\Gamma_{\mathrm{in}}$) and from the island to the absorber ($\Gamma_{\mathrm{out}}$). For the purposes of this paper, we treat the tunneling of quasiparticles in both directions as homogeneous Poisson processes, with no correlation between tunneling events and no cross-correlation between quasiparticles tunneling into and out of the SCB. At low temperatures and short time scales, this assumption has been observed both theoretically and experimentally to be invalid due to a failure of quasiparticles on the island to reach thermal equilibrium. However, at 100 mK, the proposed operation temperature of the detector, the time scales of these non-Poissonian effects are negligibly short and the approximation of uncorrelated quasiparticle tunneling is well justified.\cite{ourpaper} 

The noise spectrum of a two-rate random telegraph signal can be quickly calculated by computing the Fourier transform of the autocorrelation function.\cite{Machlup} It is found to be

\begin{equation}\label{Sapprox}
S^\mathrm{tel}_\phi (\omega) = \frac{\delta\phi^2}{\pi} 
\frac{\Gamma_\mathrm{in}\Gamma_\mathrm{out} / \Gamma_\Sigma} {\Gamma_\Sigma^2 + \omega^2}
+ \frac{\delta(\omega)}{\left( 1 + \Gamma_\mathrm{in} / \Gamma_\mathrm{out} \right)^2 }
\end{equation}

\noindent where $\Gamma_\Sigma = \Gamma_\mathrm{in} + \Gamma_\mathrm{out}$, $\Gamma_\mathrm{in} = Kn_{qp}$ and $\Gamma_\mathrm{out}$ is independent of $n_{qp}$. This is a Lorentzian-type shot noise spectral density, which consists of a flat distribution at low frequencies with a cutoff at the sum transition rate $\Gamma_\Sigma$. Note that if $\Gamma_\mathrm{in} = \Gamma_\mathrm{out}$, we recover the textbook Lorentzian expression for a single-rate random telegraph process. 

\subsection{Fano Noise}

Another key noise mechanism in the QCD is Fano noise, which arises due to the uncertainty in the number of quasiparticles generated in the absorber by an individual photon. When an incident photon is coupled to the antenna, a fast photoelectron is generated in the absorber, which rapidly thermalizes via electron-phonon collisions into an equilibrium distribution of quasiparticles with mean quasiparticle number $N_{qp} = n_{qp}\Omega$. Since this relaxation process is highly correlated, the quasiparticle generation statistics are sub-Poissonian. The variance in the quasiparticle number is not $N_{qp}$ but $FN_{qp}$, where the Fano factor $F \approx 0.2$ quantifies the degree of suppression of the fluctuations.\cite{UgoFano} In the QCD, the noise-equivalent power due to the Fano effect is given by

\begin{equation}\label{fano}
NEP_\mathrm{fano}(\omega) = \sqrt{\frac{FP_s\Delta}{\eta}\left(1+\omega^2\tau_{qp}^2\right)\left(1+\omega^2\tau_r^2\right)}.
\end{equation}

While the Fano effect is typically considered as a limitation on the energy resolution of a detector, the high sensitivity of the SCB to quasiparticle density makes this effect a small but important contribution to the overall NEP.

\subsection{Generation-Recombination Noise}

At the target QCD operating temperature of 100 mK, equilibrium quasiparticle states are strongly frozen out, so equilibrium quasiparticle tunneling is exponentially suppressed. As a result, the quasiparticle tunneling discussed above is due strictly to non-equilibrium quasiparticles. At a characteristic temperature $T^*$, equilibrium quasiparticle states begin to be activated, leading to increased tunneling rates which swamp out the response of the detector to the non-equilibrium quasiparticles generated in photon absorption. In closely related devices, $T^* \sim 250$ mK.\cite{TinkhamParity} The fluctuation in the QCD phase shift signal due to the thermal excitation and recombination of equilibrium quasiparticles is known as generation-recombination (GR) noise, with an associated frequency-independent NEP\cite{KarasikGR}

\begin{equation}\label{GRnoise}
\mathrm{NEP_{GR}} = 2\Delta \sqrt{\frac{N_L}{k_BT}} e^{-\Delta/2k_BT}
\end{equation}

\noindent where the equilibrium quasiparticle number $N_L$ is defined in Sec.~IIa. This noise arises simply from the fluctuation in quasiparticle density from equilibrium generation and recombination events, which are assumed to be uncorrelated. As shown in Fig.~6, GR noise does not impose a significant limitation on detector sensitivity at the proposed QCD operating temperature of 100 mK.

\subsection{Other Sources of Phase Noise}

In addition to telegraph noise and fluctuations in the quasiparticle density, there are a variety of physical effects which result in additional noise in the QCD phase shift signal. Both lumped-element LC oscillators and coplanar waveguide quarter-wave resonators suffer from intrinsic phase noise, which has been found to arise from coupling to a Fermionic bath of two-level charge fluctuators (TLFs) located on substrate and material surfaces.\cite{MartinisTLFs,Jiansong} This intrinsic oscillator phase noise is typically characterized by a $1/f$-type spectrum. This source of noise is currently the limiting factor for the sensitivity of MKID devices, and is also a key issue limiting the performance of phase qubits and nanomechanical oscillators. Ongoing progress in reducing this noise has focused on the investigation of novel fabrication techniques and materials, as well as improved device design. 
	
Another closely related source of phase noise in the QCD is charge noise in the SCB. This is caused by coupling of the SCB island to the bath of TLFs in the charge degree of freedom. This effectively acts as a fluctuating gate charge, which is transformed into a fluctuating phase shift in the SCB readout. This fluctuation is minimized when the SCB is operated at its charge-phase degeneracy point, although in a multiplexed array such operation is impractical with a single tuning line. 

Finally, noise in the amplifier and readout electronics will contribute additional phase noise to the QCD signal. State-of-the-art cryogenic HEMT amplifiers have a noise temperature $< 1~\mathrm{K}$ at 400-800 MHz or 5 K at 4-8 GHz.\cite{Weinreb} All of the above sources of phase noise can be measured in the QCD as additive noise on top of the telegraph signal. 

\begin{figure}[t]
\centering
\includegraphics{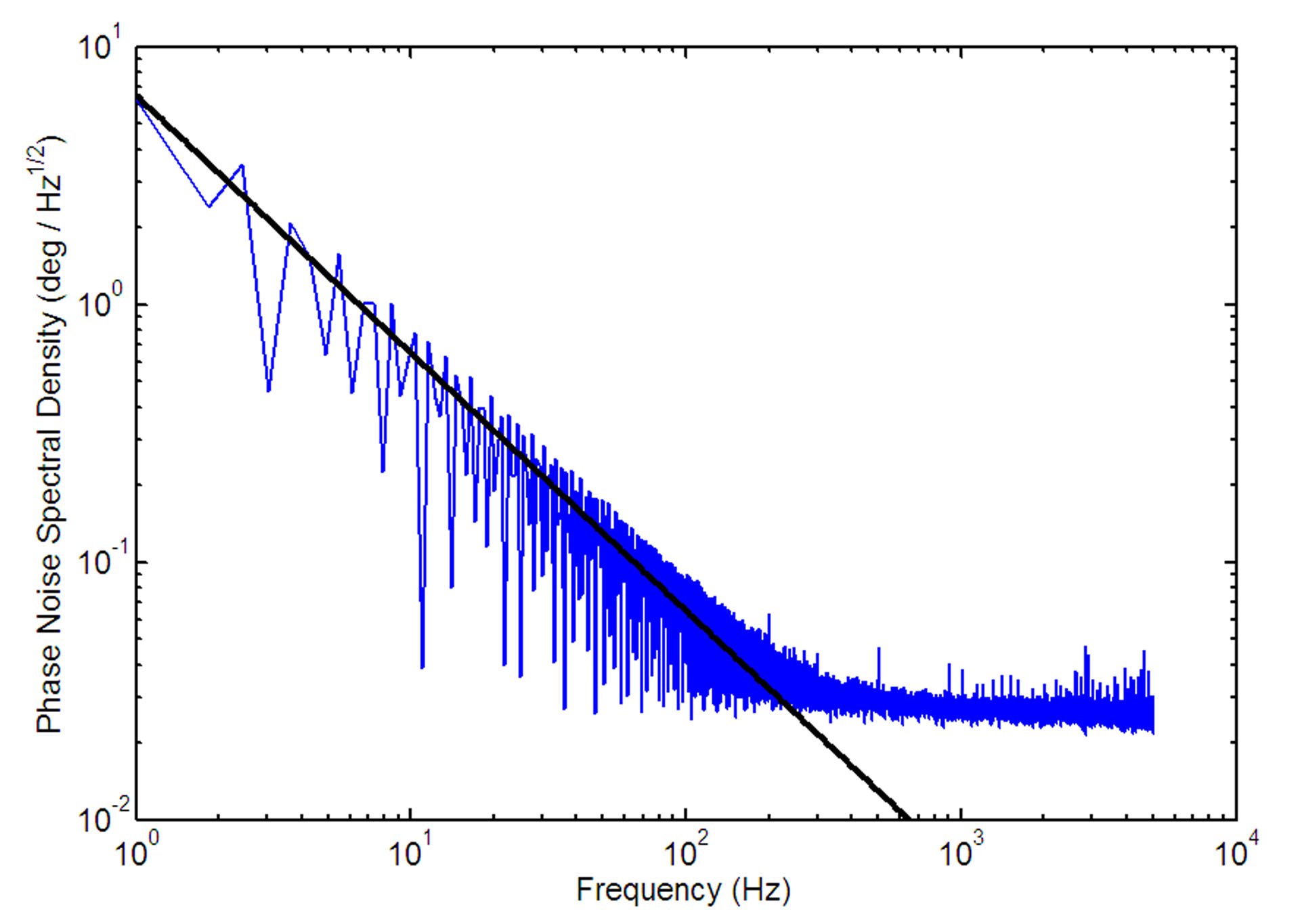}
\caption{(Color online) Measured aggregate phase noise in a test device oscillator with an input power of -125 dBm. This device was fabricated using lower-$T_C$ Al/Ti/Au trilayer leads to suppress quasiparticle tunneling. The estimated phase noise includes contributions from oscillator phase noise, SCB charge noise, and noise in the amplifiers and electronics. The blue curve is a periodogram of time-domain phase noise measured using an oscilloscope. The black line is a fit to a 1/f curve with a prefactor of 6.5 degrees at 1 Hz.}
\end{figure}

The overall aggregate phase noise, including contributions from oscillator phase noise, SCB charge noise, amplifier noise, and noise from mixers and other electronics, has been measured in a test SCB device and is shown in Fig.~4. In this device, which is not coupled to an antenna, quasiparticle tunneling is suppressed by forming the leads of the device from an Al/Ti/Au trilayer with a superconducting transition temperature of 450 mK. This creates a large energetic barrier for quasiparticle tunneling, suppressing the telegraph noise altogether. This allows us to measure the phase noise directly in the time domain, by recording data with an oscilloscope in $10^4$-point frames and computing the Fourier transform. The blue line in Fig.~4 shows a 100-average periodogram of this data, taken with a 10 kHz amplifier bandwidth. The black line is a fit to $\sqrt{S_\phi^\mathrm{phase} (\omega)} = \alpha / \omega$, where the prefactor $\alpha$ = 6.5 degrees at 1 Hz. This fit is later used with Eq.~(\ref{NEPgeneral}) to estimate the phase noise NEP in a QCD device with practical parameters.

\section{Performance Estimates}

Figures 5-12 illustrate the theoretical performance of the QCD detector using an experimentally feasible set of device parameters. In choosing detector parameters, engineering tradeoffs must be made between detector sensitivity, saturation power, and practicality of fabrication. The selected parameters were chosen as a reasonable compromise between these goals, with an eye toward applications in far-infrared and sub-millimeter astronomy. 

\begin{figure}[t]
\centering
\includegraphics{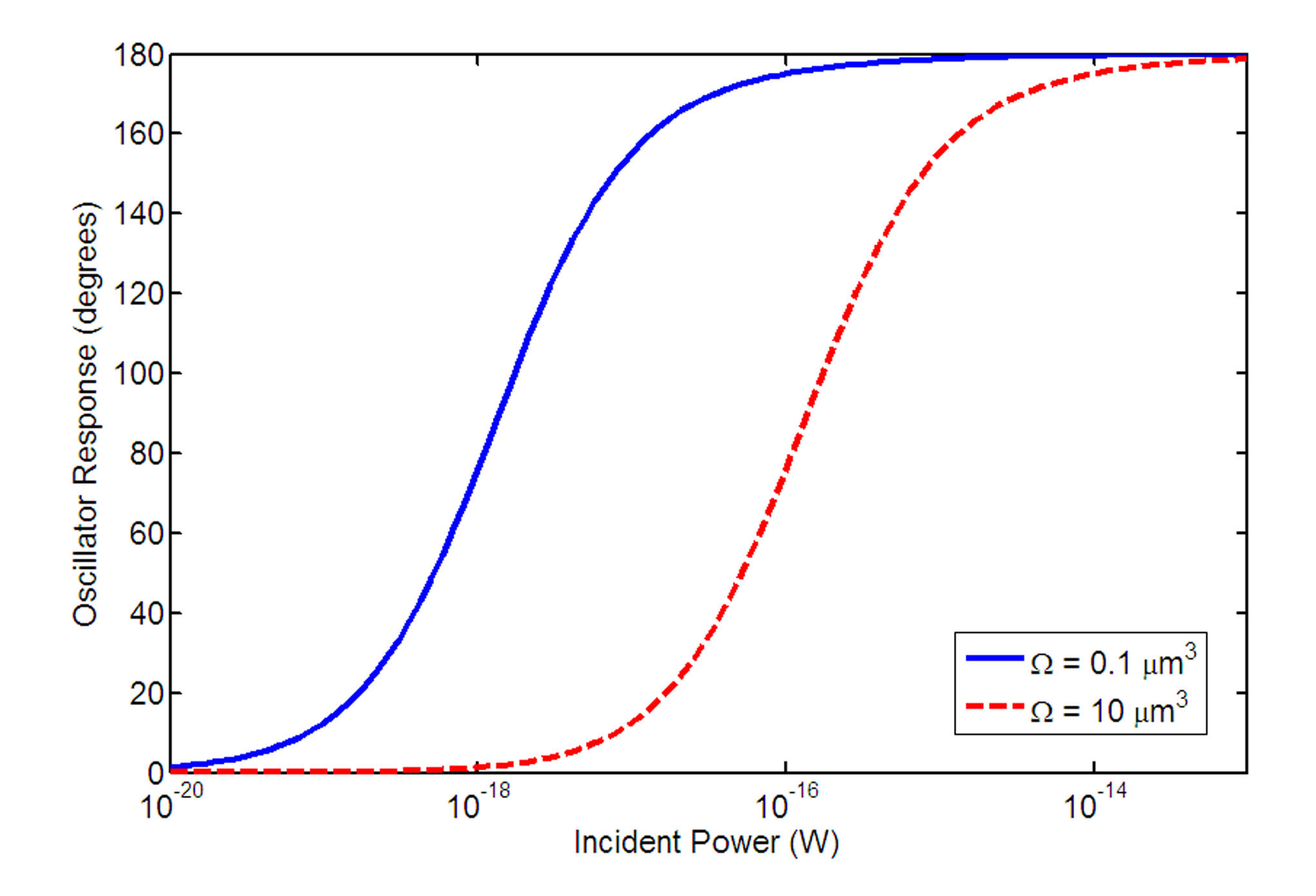}
\caption{(Color online) Theoretical average phase response of the SCB readout oscillator as a function of detector loading power. \it Solid (blue) curve:\rm~Response with small absorber, $\Omega = 0.1~\mu\mathrm{m}^3$, optimized for maximum sensitivity. \it Dashed (red) curve:\rm~Response with larger absorber, $\Omega = 10~\mu\mathrm{m}^3$, optimized for higher saturation power or shorter wavelength radiation.}
\end{figure}

Figure 5 shows a plot of the mean oscillator response $P_\mathrm{odd}\delta\phi$ as a function of incident power at the detector. This is the overall phase shift $\phi$ in the oscillator, time-averaged over the intrinsic random telegraph noise, as given by Eq.~(\ref{Peven}). The response is shown for two different absorber volumes, $\Omega = 0.1~\mathrm{\mu m}^3$ and $\Omega = 10~\mathrm{\mu m}^3$. Both curves were evaluated at a fixed temperature $T =$ 100 mK, SCB trap depth $\delta E/k_B = 0.1$ K, and tunnel barrier conductance $G_N$ = 66 $\mu\mho$. We have assumed tank circuit parameters such that the phase shift between parity states is $\delta\phi = 180^\circ$. We have also assumed the material parameters $\Delta / k_B = 2.1~\mathrm{K}$, $\tau_{qp} = 110~\mu\mathrm{s}$ and $R = 9.6~\mu\mathrm{m}^3/s$. This figure illustrates that the range of detector operation can be tailored simply by choosing an appropriate absorber volume, trading off sensitivity for saturation power. Although the response is linear over at least one decade in power, the simple functional form of Eq.~(\ref{Peven}) permits detector operation over a wider range. 

\begin{figure}[t]
\centering
\includegraphics{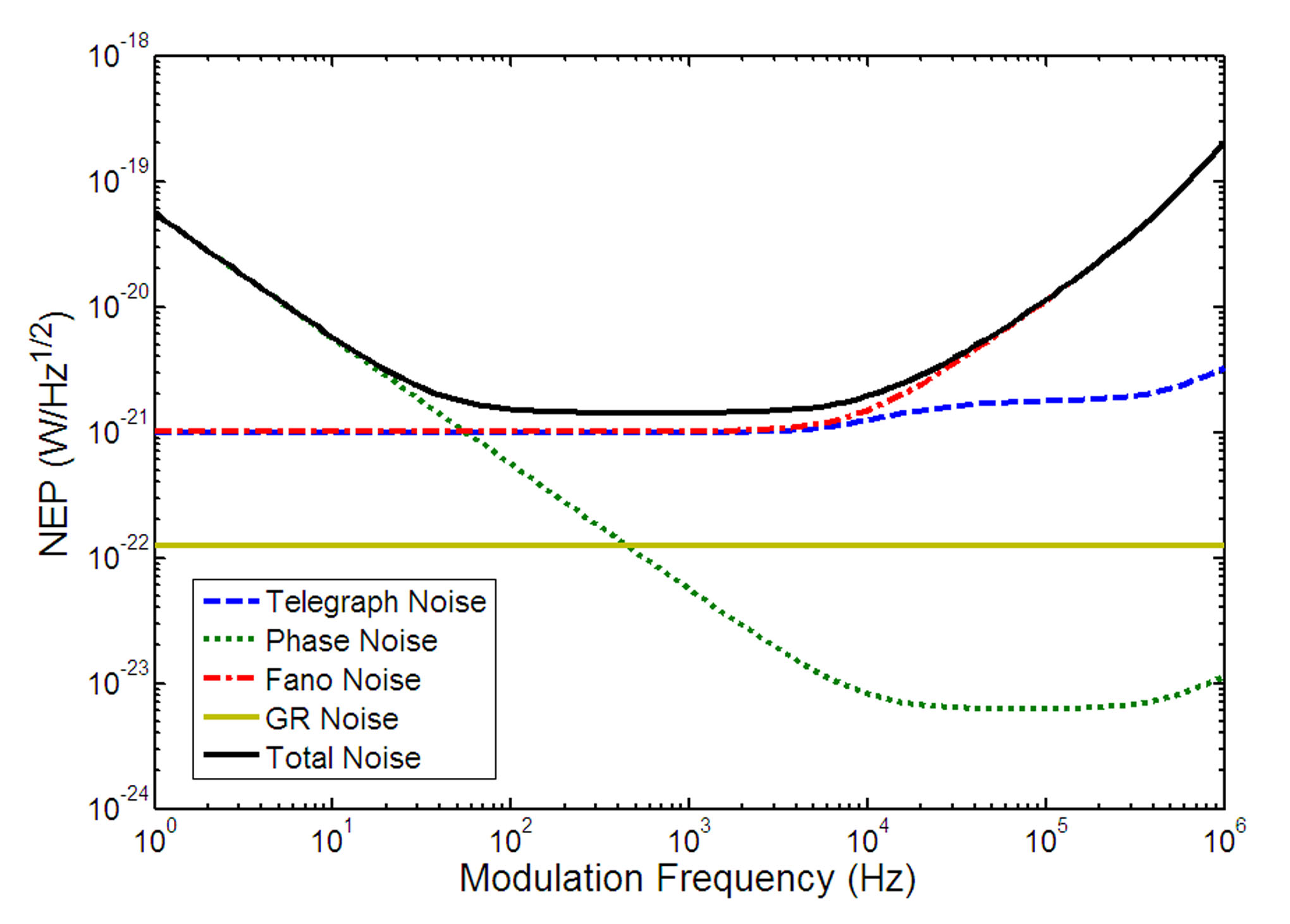}
\caption{(Color online) Noise-equivalent power as a function of signal modulation frequency for a variety of physical noise mechanisms, as described in the text. Simulations are performed with fixed temperature $T = 100$ mK and incident power $P_s = 10^{-19}$ W. \it Solid (black) line:\rm~mean-square sum of the NEP due to each noise mechanism. \it Dashed (blue) line:\rm~NEP due to telegraph noise. \it Dotted (green) line:\rm~NEP due to aggregate excess phase noise. \it Dash-dotted (red) line:\rm~NEP due to Fano noise. \it Solid horizontal (gold) line:\rm~NEP due to generation-recombination noise.}
\end{figure}

Figure 6 shows the theoretical sensitivity of the detector as a function of signal modulation frequency for the design parameters used to compute Fig.~5. The absorber volume used was $\Omega = 0.1~\mathrm{\mu m}^3$, which is the design optimized for maximum sensitivity. The incident power was taken to be $P_s = 10^{-19}$ W. The plot illustrates NEP due to the four dominant noise mechanisms described in section III as a function of signal modulation frequency. Since the noise arising from different physical mechanisms is uncorrelated, the solid (black) line is the mean-square sum of the individual NEP values for each mechanism. As can be seen from Eq.~(\ref{NEPgeneral}), the functional form of each NEP curve is dominated by the spectral density of noise at low frequencies, and is ultimately limited at high frequencies by the quasiparticle lifetime and the response time of the readout oscillator. 

The dashed (blue) line is the NEP due to telegraph noise, as found by substituting Eq.~(\ref{Sapprox}) into Eq.~(\ref{NEPgeneral}). The shape of this curve is dominated by the Lorentzian form of the noise spectral density, which is flat at low modulation frequencies and falls off sharply at the effective tunneling rate $\Gamma_\Sigma$. The particular value of $\Gamma_\Sigma = 26~\mathrm{kHz}$ shown in Fig.~6 emerges from the particular choice of the incident power $P_s$, and will change with optical loading in keeping with Eq.~(\ref{Nqp}). For this particular set of device parameters, $\Gamma_\mathrm{out} = 15$ kHz. It is important to note that since the observed signal in the QCD is the phase shift in the reflected wave averaged over the telegraph noise, it is meaningless to operate the detector at a modulation frequency faster than the cutoff. As a result, operation must be restricted to a modulation frequency $1/\tau_m \ll \Gamma_{\mathrm{out}}$. For the remainder of the calculations in this paper, we have taken the operation time to be $\tau_m = 10~\mathrm{ms}$. 

The dotted (green) line is the NEP due to the aggregate excess phase noise as a function of frequency, adapted from the measurements shown in Fig.~4. The plot is of Eq.~(\ref{NEPgeneral}) with $\sqrt{S_\phi (\omega)} = \alpha / \omega$ where $\alpha$ = 6.5 degrees at 1 Hz. Note that this noise source is dominant at low modulation frequencies, where the detector is likely to be operated. However, the magnitude of this noise is not currently at a fundamental limit, and can likely be improved by using better amplifiers and electronics, along with lower-noise substrates and materials. This will significantly improve the sensitivity of the detector at low frequencies. The dash-dotted (red) line is the NEP due to the Fano noise, as estimated by Eq.~(\ref{fano}). Note that for the small absorber volume considered here, the Fano noise is actually equal to or in excess of the telegraph noise. Finally, the solid horizontal (gold) line is the NEP due to generation-recombination noise, which in the approximation of Eq.~(\ref{GRnoise}) is taken to be frequency-independent. At 100 mK, the design operation temperature of the device, equilibrium quasiparticle tunneling is strongly frozen out and the GR noise is a negligible contribution to the total. 
 
\begin{figure}[t]
\centering
\includegraphics{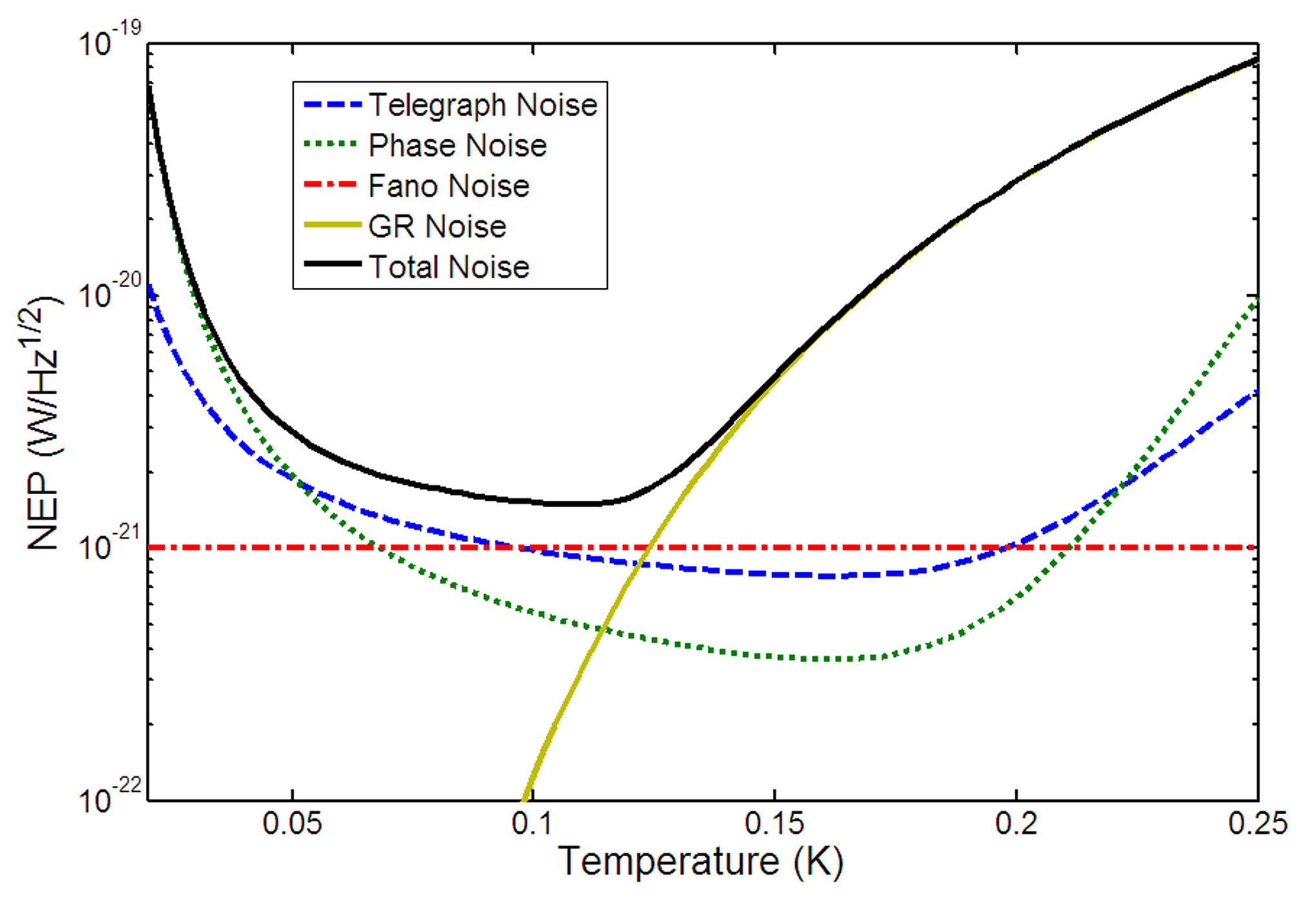}
\caption{(Color online) Noise-equivalent power as a function of operating temperature for a variety of physical noise mechanisms, as described in the text. Simulations are performed with fixed measurement time $\tau_m = 10~\mathrm{ms}$ and incident power $P_s = 10^{-19}$ W. The color scheme and line styles are identical to those of Fig.~6.}
\end{figure}

Figure 7 shows the theoretically estimated NEP as a function of operating temperature, for the same physical noise mechanisms discussed in Fig.~6. In this case, the operating time $\tau_m = 10~\mathrm{ms}$, $\Omega = 0.1~\mathrm{\mu m}^3$, and all other parameters are the same as in figs.~5 and 6. The color scheme for the plots is identical to that of Fig.~6 for NEP due to telegraph noise, excess phase noise, Fano noise, and generation-recombination noise. At very low temperatures, it is interesting to observe that the NEP increases sharply. This is due to the sharp increase in the island-to-absorber tunneling rate $\Gamma_\mathrm{out}$ which occurs at low temperatures, where quasiparticles in the SCB island fail to reach thermal equilibrium and tunnel elastically.\cite{ourpaper} This sharply decreases the responsivity of the detector at low temperature, increasing the NEP. Also note that in the proposed region of operation, 100 mK, the sensitivity of the detector is quite flat. Above 150 mK, tunneling of quasiparticles in thermal equilibrium begins to dominate, and the generation-recombination noise becomes the dominant noise source. Note that the Fano noise is independent of temperature, since the temperature dependence enters into the NEP through the responsivity, which cancels in the derivation of Eq.~(\ref{fano}).

Figure 8 shows the detector NEP as a function of loading power, calculated using the same expressions as Figs.~6 and 7. The design parameters are the same as those used in Fig.~7, with the temperature held at 100 mK and $\Omega = 0.1~\mathrm{\mu m}^3$. The solid (blue) line is the total NEP, which is the mean-square sum of the NEPs from telegraph noise, excess phase noise, Fano noise, and GR noise. Note that the NEP increases dramatically as the incident power approaches $10^{-16}$ W. At this power level, the detector is approaching saturation, as can be seen from Fig.~5. 

For far-infrared and submillimeter radiation, another key qualification is whether the detector is sensitive enough to be limited by the shot noise of the incident signal itself. To compare the QCD sensitivity with the shot noise limit, we have plotted the NEP due to photon shot noise $\mathrm{NEP}_\gamma = \sqrt{2 P_s \hbar c / \lambda}$ at two wavelengths, $\lambda = 30~\mu\mathrm{m}$ and $\lambda = 1$ mm, as shown by the dashed (green) and dotted (red) lines, respectively. From this figure, one can see that the QCD detector is predicted to be shot-noise limited over a wide range of operating powers. 

In the same vein, Fig.~9 shows a plot of the saturation power of the detector as a function of frequency compared to the shot noise limits. In this plot, the solid (blue) curve is the saturation power $P_\mathrm{sat}$, defined as the power at which the response $\phi$ of the detector falls within one noise standard deviation $\sigma = \sqrt{S_\phi(\omega) \omega}$ of the phase response limit $\delta\phi$, evaluated at a frequency $\omega$. The overall structure follows the frequency dependence shown in Fig.~6, with saturation power decreasing at low frequencies due to $1/f$ phase noise at at high frequencies due to the finite quasiparticle lifetime and finite response time of the readout oscillator. The dashed (green) and dotted (red) lines are an illustration of the critical power above which the detector ceases to be shot-noise limited, as illustrated in Fig.~8. The points where the shot noise limits fall rapidly indicate frequency ranges over which the detector is not shot-noise limited at all. The dashed and dotted curves in Fig.~9 correspond to the shot noise limits for $\lambda = 30~\mu\mathrm{m}$ and $\lambda = 1$ mm, respectively.

\begin{figure}[t]
\centering
\includegraphics{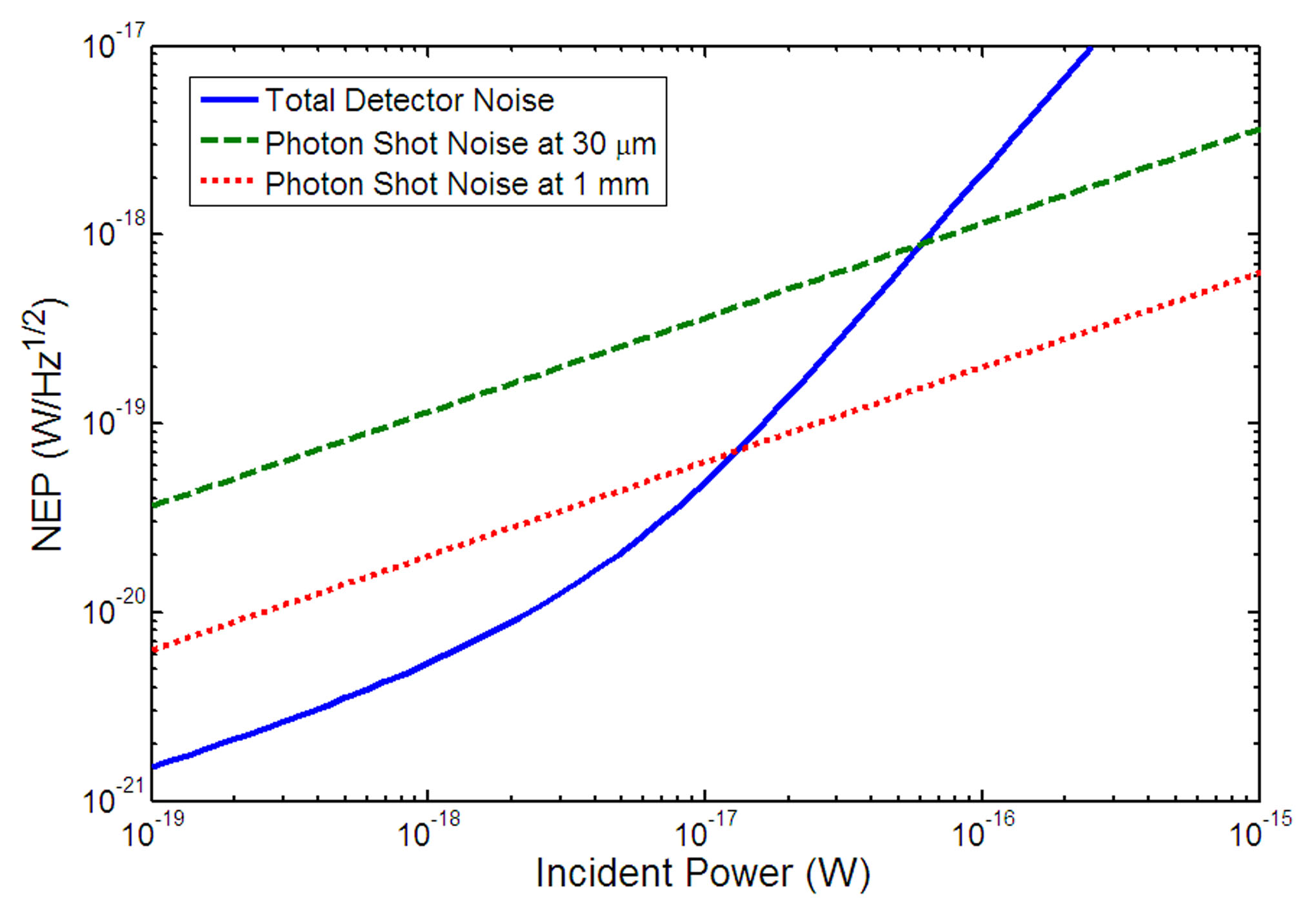}
\caption{(Color online) Noise-equivalent power as a function of loading power, as described in the text. Simulations are performed with fixed modulation time $\tau_m = 10~\mathrm{ms}$ and operating temperature $T = 100~\mathrm{mK}$. \it Solid (blue) line:\rm~mean-square total NEP due to all noise mechanisms. \it Dashed (green) line:\rm~Photon shot noise at a wavelength $\lambda = 30~\mu\mathrm{m}$. \it Dotted (red) line:\rm~Photon shot noise at $\lambda = 1~\mathrm{mm}.$}
\end{figure}

\begin{figure}[t]
\centering
\includegraphics{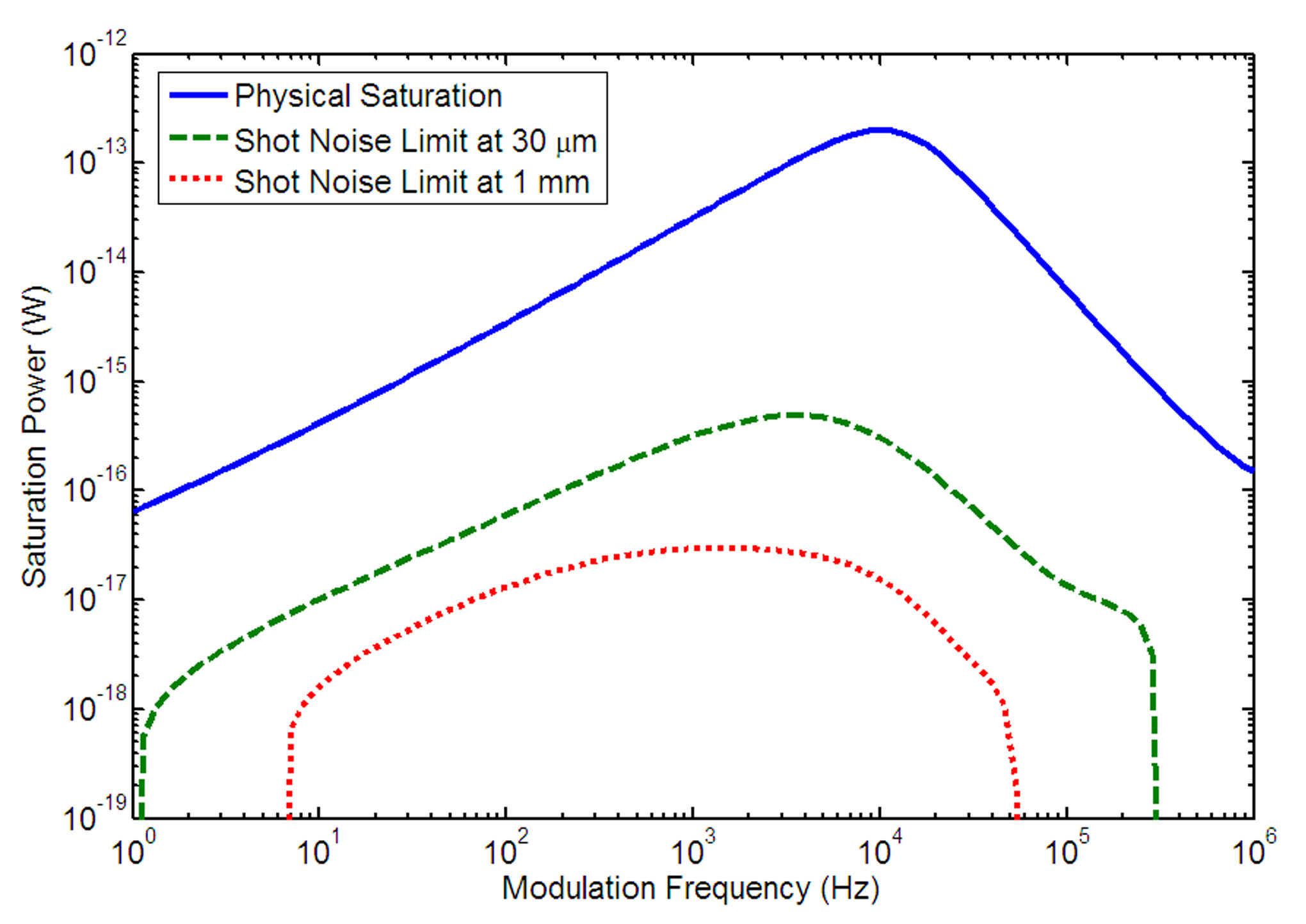}
\caption{(Color online) Saturation power as a function of frequency, for the sample parameters used in Fig.~8. \it Solid (blue) curve:\rm~Physical saturation power of the detector, as defined in the text. \it Dashed (green) curve:\rm~Critical power at which the detector ceases to be shot noise limited, as defined in the text, for $\lambda = 30~\mu m$. \it Dotted (red) curve:\rm~Shot noise limit for $\lambda = 1$ mm. The points where the shot noise limits drop rapidly off scale indicate frequency ranges where the detector is not shot noise limited at all.}
\end{figure}

\begin{figure}[t]
\centering
\includegraphics{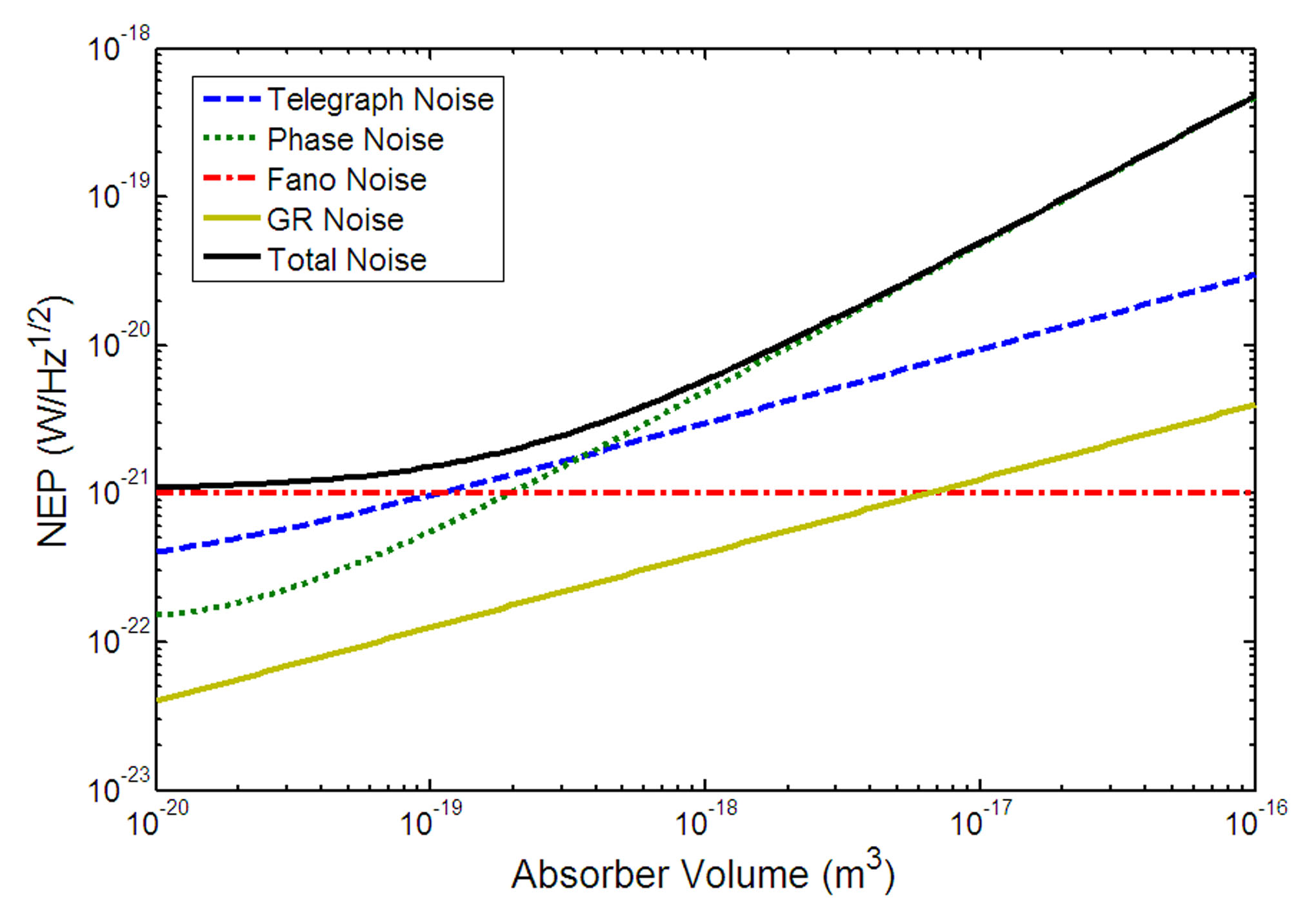}
\caption{(Color online) Noise-equivalent power as a function of absorber volume $\Omega$, for fixed measurement time $\tau_m = 10~\mathrm{ms}$, operating temperature $T = 100~\mathrm{mK}$, and incident power $P_s = 10^{-19}$ W. The color scheme and line styles are identical to those of Figs.~6 and 7. While the sensitivity of the detector improves with a smaller absorber, engineering tradeoffs must be made with saturation power and practicality of fabrication.}
\end{figure}

As we have stressed above, a critical QCD design parameter is the absorber volume $\Omega$. Fig.~10 shows theoretical NEP as a function of $\Omega$ for the four noise sources considered above. The colors and line styles for the plot are the same as for Figs.~6 and 7. As can be seen from Fig.~5, the sensitivity of the device improves as the absorber volume is reduced, assuming that the absorber-island tunneling rate depends linearly on the quasiparticle density. This improvement in sensitivity clearly comes at the expense of saturation power, so $\Omega$ must be tailored for a specific application. The absorber cannot be made arbitrarily small, since it must be coupled to the antenna and must be comparable in volume to the SCB island. Furthermore, the absorber must be large enough for the generated quasiparticle population to equilibrate below the gap edge of the Nb antenna within the diffusion time, to ensure quasiparticle confinement in the absorber region. In the calculations discussed above, we have set $\Omega = 10^{-19}~\mathrm{m}^3$, which is a reasonable compromise between these concerns. As with the temperature dependence, note that the Fano noise is volume-independent.

\section{Applications}

Such sensitive submillimeter radiation detectors are critically needed for future experiments in far-infrared astrophysics. Approximately half of the total light emitted from stars and black hole accretion over the history of the universe has been absorbed by dust and reradiated in the 10-1000 $\mu\mathrm{m}$ band. However, detailed spectroscopic investigations of this spectral region have remained difficult due to the challenges of ground-based observation and a need for sensitive detectors. While significant advances have recently been made with germanium photoconductors and beam-isolated silicon nitride TES bolometers, the QCD could play an important role as a sensitive detector which can be frequency-multiplexed into large arrays in a straightforward fashion. 

\begin{figure}
\centering
\includegraphics{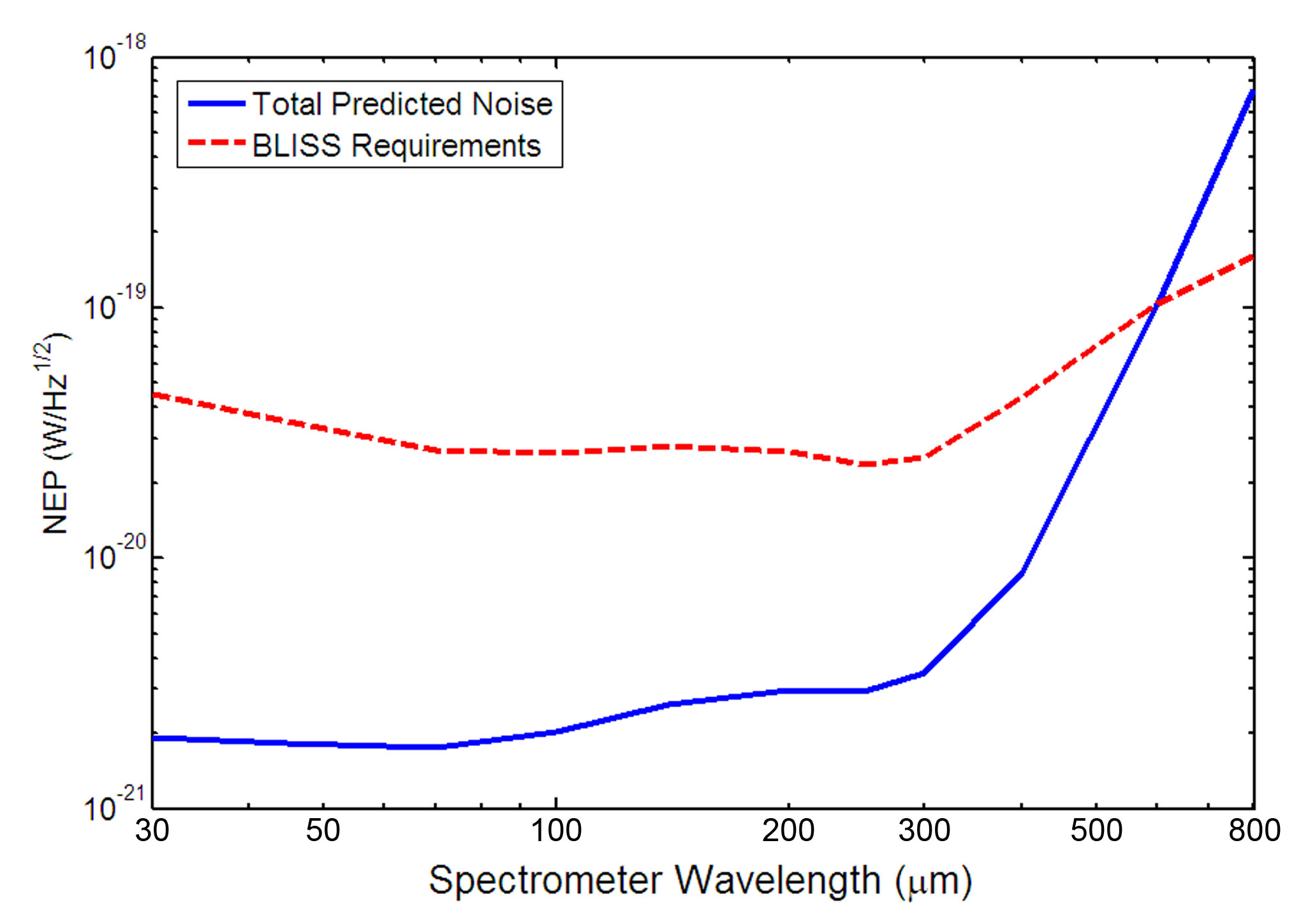}
\caption{(Color online) Comparison of QCD detector sensitivity as a function of wavelength with the detector requirements for the BLISS spectrometer. Detector requirements and spectrometer loading data are adapted from Ref.~15.}
\end{figure}

A proposed instrument which is a natural fit for the QCD detector is the Background-Limited Infrared-Submillimeter Spectrometer
(BLISS), a spectrograph designed for the Space Infrared Telescope for Cosmology and Astrophysics (SPICA).\cite{BLISS} SPICA is a 3.5 m, 4.5 K cold space telescope planned for launch in 2013. One of the key technological challenges for realizing the BLISS spectrometer is the development of a sensitive submillimeter detector array. In Fig.~12, we show the detector sensitivity requirements for the BLISS spectrometer alongside the theoretical NEP for the QCD detector, plotted as a function of incident radiation wavelength. For each wavelength, we have taken the design optical loading power for the BLISS spectrometer and computed the theoretical NEP as in Fig.~8, using the QCD design parameters given above. For most of the BLISS spectral region, the QCD detector meets these sensitivity requirements by a wide margin. The sharp increase in NEP at long wavelengths is due to increased optical loading in the spectrometer design. At $\lambda = 30~\mu\mathrm{m}$, the design loading power is $P_s = 1.62 \times 10^{-19}$ W, while at $\lambda = 800~\mu\mathrm{m}$, the loading power $P_s = 5.4 \times 10^{-17}$ W. 

Because the QCD sensitivity can be tuned to the quasiparticle density by varying the absorber volume, the QCD may also be useful in applications at shorter wavelengths. Furthermore, there exist a wide variety of non-astrophysical applications for sensitive detectors in far-infrared and submillimeter regime, including earth science, planetary science, biomedical technology and defense applications. 

\section{Conclusions}

We describe a scheme for a sensitive pair-breaking photon detector applicable for far-infrared and sub-millimeter radiation. Incident photons coupled with an antenna break Cooper-pairs in a superconducting absorber. A single Cooper-pair box in contact with the absorber is used as a probe to measure the density of quasiparticles. The SCB signal can be read out by using a radio-frequency capacitance measurement or by embedding the SCB in a microwave resonator. This scheme lends itself naturally to large-scale frequency multiplexing. The sensitivity and performance of the proposed device has been calculated, including several physical noise sources. These sources include telegraph noise, fano noise, generation-recombination noise, and excess resonator, charge and electronic phase noise. The detector has very favorable predicted sensitivities, with a minimum NEP on the order of $10^{-21}~\mathrm{W}/\sqrt{\mathrm{Hz}}$, and can be readily fabricated using existing techniques. The QCD promises excellent sensitivity, and is a natural tool for applications in far-infrared astrophysics. 

\acknowledgements{We would like to thank Richard Muller for performing the electron-beam lithography, and Jonas Zmuidzinas,  Per Delsing, and Sunil Golwala for helpful discussions and advice. This work was conducted at the Jet Propulsion Laboratory, California Institute of Technology, under a contract with the National Aeronautics and Space Administration (NASA). Matthew Shaw acknowledges support from the University of Southern California College of Arts and Sciences. Juan Bueno acknowledges support from NASA. Government sponsorship acknowledged.}


\begin{thebibliography}{1}

\bibitem{Peacock} A. Peacock, P. Verhoeve, N. Rando, A. van Dordrecht, B. G. Taylor, C. Erd, M. A. C. Perryman, R. Venn, J. Howlett, D. J. Goldie, J. Lumley, and M. Wallis, Nature \bf 381\rm, 135 (1996)

\bibitem{particles} N. E. Booth and D. J. Goldie, Supercond. Sci. Tech \bf 9\rm, 493 (1996)

\bibitem{saewoo} A. J. Miller, S. W. Nam, J. M. Martinis, and A. V. Sergienko, Appl. Phys. Lett \bf 83\rm, 791 (2003)

\bibitem{nanowire} B. S. Robinson, A. J. Kerman, E. A. Dauler, R. J. Barron, D. O. Caplan, M. L. Stevens, J. J. Carney, and S. A. Hamilton, Opt. Lett. \bf 31\rm, 444 (2006)

\bibitem{firstTES} K. D. Irwin, G. C. Hilton, D. A. Wollman, and J. M. Martinis, Appl. Phys. Lett. \bf 69\rm, 1945 (1996)

\bibitem{firstHEB} D. E. Prober, Appl. Phys. Lett. \bf 62\rm, 2119 (1993)

\bibitem{nanoHEB} J. Wei, D. Olaya, B. S. Karasik, S. V. Pereverzev, A. V. Sergeev, and M. E. Gershenson, Nat. Nano. \bf 3\rm, 496 (2008)

\bibitem{firstSTJ} G. H. Wood and B. L. White, Appl. Phys. Lett. \bf 15\rm, 237 (1969)

\bibitem{MKID} P. K. Day, H. G. LeDuc, B. A. Mazin, A. Vayonakis, and J. Zmuidzinas, Nature \bf 425\rm, 817 (2003).

\bibitem{schonRMP} Y. Makhlin, G. Sch\"{o}n, and A. Shnirman, Rev. Mod. Phys. \bf 73\rm, 357 (2001)

\bibitem{Aumentado1} J. Aumentado, M. W. Keller, J. M. Martinis, and M. H. Devoret, Phys. Rev. Lett \bf 92\rm, 066802 (2004)

\bibitem{Aumentado2} O. Naaman and J. Aumentado, Phys. Rev. B \bf 73\rm, 172504 (2006)

\bibitem{Ferguson} A. J. Ferguson, N. A. Court, F. E. Hudson, and R. G. Clark, Phys. Rev. Lett. \bf 97\rm, 106603 (2006)

\bibitem{ourpaper} M. D. Shaw, R. M. Lutchyn, P. Delsing, and P. M. Echternach, Phys. Rev. B \bf 78\rm, 024503 (2008)

\bibitem{BLISS} C. M. Bradford and T. Nakagawa, New Astro. Rev. \bf 50\rm, 221 (2006)

\bibitem{DelsingQCR} T. Duty, G. Johansson, K. Bladh, D. Gunnarsson, C. Wilson, and P. Delsing, Phys. Rev. Lett. \bf 95\rm, 206807 (2005)

\bibitem{FinnQCR} M. A. Sillanp\"{a}\"{a}, T. Lehtinen, A. Paila, Yu. Makhlin, L. Roschier, and P. J. Hakonen, Phys. Rev. Lett. \bf 95\rm, 206806 (2005). 

\bibitem{BornQCR} D. Born, V. I. Shnyrkov, W. Krech, Th. Wagner, E. Il'ichev, M. Grajcar, U. H\"{u}bner, and H.-G. Meyer, Phys. Rev. B \bf 70\rm, 180501(R) (2004)

\bibitem{tinkham} J. M. Hergenrother, J. G. Lu, M. T. Tuominen, D. C. Ralph, and M. Tinkham, Phys. Rev. B \bf 51\rm, 9407 (1995)

\bibitem{schoelkopf_detector} R. J. Schoelkopf, S. H. Moseley, C. M. Stahle, P. Wahlgren, and P. Delsing, IEEE Trans. Appl. Sup. \bf 9\rm, 2935 (1999)

\bibitem{MKIDcamera} J. Glenn, P. K. Day, M. Ferry, J. Gao, S. R. Golwala, S. Kumar, H. G. LeDuc, P. R. Maloney, B. A. Mazin, H. Nguyen, O. Noroozian, J. Savers, J. Schlaerth, J. E. Villancourt, A. Vayonakis, and J. Zmuidzinas, Proc. SPIE \bf 7020\rm, 70200B (2008)

\bibitem{ourMultiplexed} M. D. Shaw, J. F. Schneiderman, J. Bueno, B. S. Palmer, P. Delsing, and P. M. Echternach, Phys. Rev. B \bf 79\rm, 014516 (2009)

\bibitem{Lutchyn} R. M. Lutchyn and L. I. Glazman, Phys. Rev. B \bf 75\rm, 184520 (2007)

\bibitem{nakamura_filmthickness} T. Yamamoto, Y. Nakamura, Yu. A. Pashkin, O. Astafiev, and J. S. Tsai, Appl. Phys. Lett. \bf 88\rm, 212509 (2006) 

\bibitem{ferguson_filmthickness} N. A. Court, A. J. Ferguson, and R. G. Clark, Supercond. Sci. Tech. \bf 21\rm, 015013 (2008)

\bibitem{QPdownconversion} A. G. Kozorezov, A. F. Volkov, J. K. Wigmore, A. Peacock, A. Poelaert, and R. den Hartog, Phys. Rev. B \bf 61\rm, 11807 (2000)

\bibitem{Gaidis} M. C. Gaidis, Ph.D. Thesis, Yale University (1994)

\bibitem{Machlup} S. Machlup, J. Appl. Phys. \bf 25\rm, 341 (1954)

\bibitem{UgoFano} U. Fano, Phys. Rev. \bf 72\rm, 26 (1947)

\bibitem{TinkhamParity} M. T. Tuominen, J. M. Hergenrother, T. S. Tighe, and M. Tinkham, Phys. Rev. Lett. \bf 69\rm, 1997 (1992)

\bibitem{KarasikGR} A. V. Sergeev, V. V. Mitin, and B. S. Karasik, Appl. Phys. Lett. \bf 80\rm, 817 (2002)

\bibitem{MartinisTLFs} J. M. Martinis, K. B. Cooper, R. McDermott, M. Steffen, M. Ansmann, K. D. Osborn, K. Cicak, S. Oh, D. P. Pappas, R. W. Simmonds, and C. C. Yu, Phys. Rev. Lett. \bf 95\rm, 210503 (2005)

\bibitem{Jiansong} J. Gao, M. Daal, A. Vayonakis, S. Kumar, J. Zmuidzinas, B. Sadoulet, B. A. Mazin, P. K. Day, and H. G. LeDuc, Appl. Phys. Lett. \bf 92\rm, 152505 (2008)

\bibitem{Weinreb} S. Weinreb, J. C. Bardin, and H. Mani, IEEE Trans. Microwave Theory and Tech. \bf 55\rm, 2306 (2007)

\end{thebibliography}
\end{document}